\documentclass[10pt,conference,a4paper]{IEEEtran}
\IEEEoverridecommandlockouts
\usepackage{cite}
\usepackage{amsmath,amssymb,amsfonts}
\usepackage{algorithmic}
\usepackage{graphicx}
\usepackage{textcomp}
\usepackage{xcolor}
\usepackage{multirow}
\def\BibTeX{{\rm B\kern-.05em{\sc i\kern-.025em b}\kern-.08em
    T\kern-.1667em\lower.7ex\hbox{E}\kern-.125emX}}

\begin{document}

\title{\fontsize{22pt}{24pt}\selectfont RadioPiT: Radio Map Generation with Pixel Transformer Driven by Ultra-Sparse Real-World Data\\
}

\author{\IEEEauthorblockN{Zeyao Sun\IEEEauthorrefmark{1}, Bohao Fan\IEEEauthorrefmark{1}, Qingyu Liu\IEEEauthorrefmark{1},
Shuhang Zhang\IEEEauthorrefmark{2},
and Lingyang Song\IEEEauthorrefmark{1}\IEEEauthorrefmark{2}\IEEEauthorrefmark{3}}
\IEEEauthorblockA{\IEEEauthorrefmark{1}School of Electronic and Computer Engineering, Peking University Shenzhen Graduate School, Shenzhen, China}
\IEEEauthorblockA{\IEEEauthorrefmark{2}School of Electronics, Peking University, Beijing, China}
\IEEEauthorblockA{\IEEEauthorrefmark{3}Hunan Institute of Advanced Sensing and Information Technology, Xiangtan University, Xiangtan, China}
}

\maketitle

\begin{abstract}
As wireless communication networks rapidly evolve, spectrum resources are increasingly scarce, making effective spectrum management critically important. Radio map is a spatial representation of signal characteristics across different locations in a given area, which serves as a key tool for enabling precise spectrum management. To generate accurate radio maps, extensive research efforts have been made. However, most existing studies are conducted on simulation data, which differs significantly from real-world data and cannot accurately reflect the spectrum characteristics of practical environments. To tackle this problem, we construct a dataset of real-world radio map with a self-developed measurement system. Due to the limited volume of real-world data and the distributional discrepancies between simulation and real-world data, we propose a Pixel Transformer (PiT)-based model enhanced with the test-time adaptation (TTA) strategy, named RadioPiT, for real-world radio map generation. Experimental results demonstrate that our proposed RadioPiT significantly outperforms baseline methods in real-world scenarios, yielding a 21.9\% decrement in the root mean square error (RMSE) compared to RadioUNet.
\end{abstract}

\begin{IEEEkeywords}
Radio map generation, real-world data, Pixel Transformer, test-time adaptation
\end{IEEEkeywords}

\section{Introduction}
 A radio map~\cite{1} is a spatial representation of signal characteristics such as received signal strength or path loss across a geographic area. In scenarios such as smart cities~\cite{2}, unmanned aerial vehicle communications~\cite{3},~\cite{4}, and vehicular networks~\cite{5}, the use of radio maps has become increasingly dense for spectrum management. However, due to the constraint of measurement costs, only ultra-sparse radio data can be collected. To address this limitation, Generative Artificial Intelligence (GAI)~\cite{6} has emerged as a powerful tool, capable of learning complex spatial patterns and generating high-fidelity outputs from limited data. Accordingly, the adoption of GAI techniques~\cite{7} is crucial for constructing fine-grained radio maps with ultra-sparse radio data.

Recently, some initial works on radio map generation with GAI-based methods have been studied. Levie et al.~\cite{8} proposed RadioUNet, a radio map estimation approach that employs convolutional neural networks (CNNs) to learn the mapping between urban environment features and path loss distributions. Zhang et al.~\cite{9} developed a cooperative radio map estimation framework, termed GAN-CRME, which leveraged a conditional generative adversarial network (GAN) to infer the spatial signal distribution from distributed received signal strength measurements and geographical maps. Shao et al.~\cite{10} designed RobUNet, a UNet-based radio map construction method designed to achieve strong generalization capability under unknown system conditions. However, these studies are all conducted on simulation data, which exhibits substantial differences from real-world data.

To facilitate radio map research on real-world data, we build a hardware-software integrated system to collect spectrum data and construct a real-world radio map dataset. The dataset covers measurements from three typical types of areas—urban, suburban, and campus environments—and includes five frequency bands with concentrated environmental signals. In total, 4,500 spectrum measurements are collected, providing a realistic foundation for evaluating radio map generation methods in real-world scenarios.

However, adopting the real-world dataset for radio map generation model training faces several challenges. First, unlike simulation radio maps that provide signal values continuously across all pixels, real-world radio maps contain measured ground-truth values only at discrete, ultra-sparse pixels, posing significant challenges for model training. Second, due to the high cost of data collection, the scale of the real-world radio map dataset is limited and insufficient to support model training independently. Third, there exists significant distributional discrepancies between the simulation and real-world data. As a result, models pre-trained on the simulation radio map dataset exhibit notable drops in radio map generation accuracy when applied to real-world data. It is therefore necessary to adopt strategies that ensure the model maintains robust performance on real-world data.

To address the above challenges, we propose a RadioPiT model based on Pixel Transformer (PiT)~\cite{11}, which supports per-pixel generation for real-world radio maps. This enables the model to cope with the difficulty posed by the discrete availability of ground-truth values in the real-world radio map dataset. We first pre-train the RadioPiT model on a simulation radio map dataset to learn the prior distribution of signal propagation. To mitigate the performance degradation caused by the distributional discrepancies between simulation and real-world data, we then propose to incorporate a test-time adaptation (TTA)~\cite{12},~\cite{13} strategy into the testing phase of the RadioPiT model, enabling online fine-tuning of model parameters. This further improves the accuracy of radio map generation on real-world data.

The main contributions of this paper are as follows:

\begin{itemize}
\item \textbf{Dataset Construction}: We build a hardware-software integrated system to construct a real-world radio map dataset with 4,500 spectrum measurements collected from three representative areas (urban, suburban, and campus), covering five concentrated environmental frequency bands.

\item \textbf{Model Proposal}: We propose the RadioPiT model based on the PiT and the TTA strategy, which accurately predicts signal strength values for radio map generation by employing the attention-enabled per-pixel generation mechanism and mitigating the impact of distributional discrepancies between simulation and real-world data.

\item \textbf{Performance Evaluation}: Experiments demonstrate that our proposed RadioPiT significantly outperforms baseline methods in terms of real-world radio map generation, yielding a 21.9\% decrement in the root mean square error (RMSE) compared to RadioUNet.
\end{itemize}

The remainder of this paper is organized as follows. In Section II, we describe the construction of the real-world radio map dataset. In Section III, we present the model design of RadioPiT. In Section IV, we perform radio map generation experiments to study the performance of RadioPiT. Section V concludes the paper.

\section{Real-world Radio Map Dataset}

In this section, we describe the real-world radio map dataset constructed from manually collected real-world data.

\subsection{Hardware-Software Integrated System Setup}

As shown in Fig. 1, the data collection hardware comprises a spectrum analyzer (Nijia Electronic Technology SA-60), an antenna, a laptop, a high-capacity portable power bank, and a data collection vehicle. The software system consists of the spectrum analyzer’s companion application (FSStudio) and a map-based calibration tool. FSStudio performs real-time acquisition and logging of signal strength across frequency bands, while the calibration software records the geographic positions of each measurement and generates the corresponding latitude–longitude coordinate files.

\begin{figure}[h]
    \centering
    \includegraphics[width=1\linewidth]{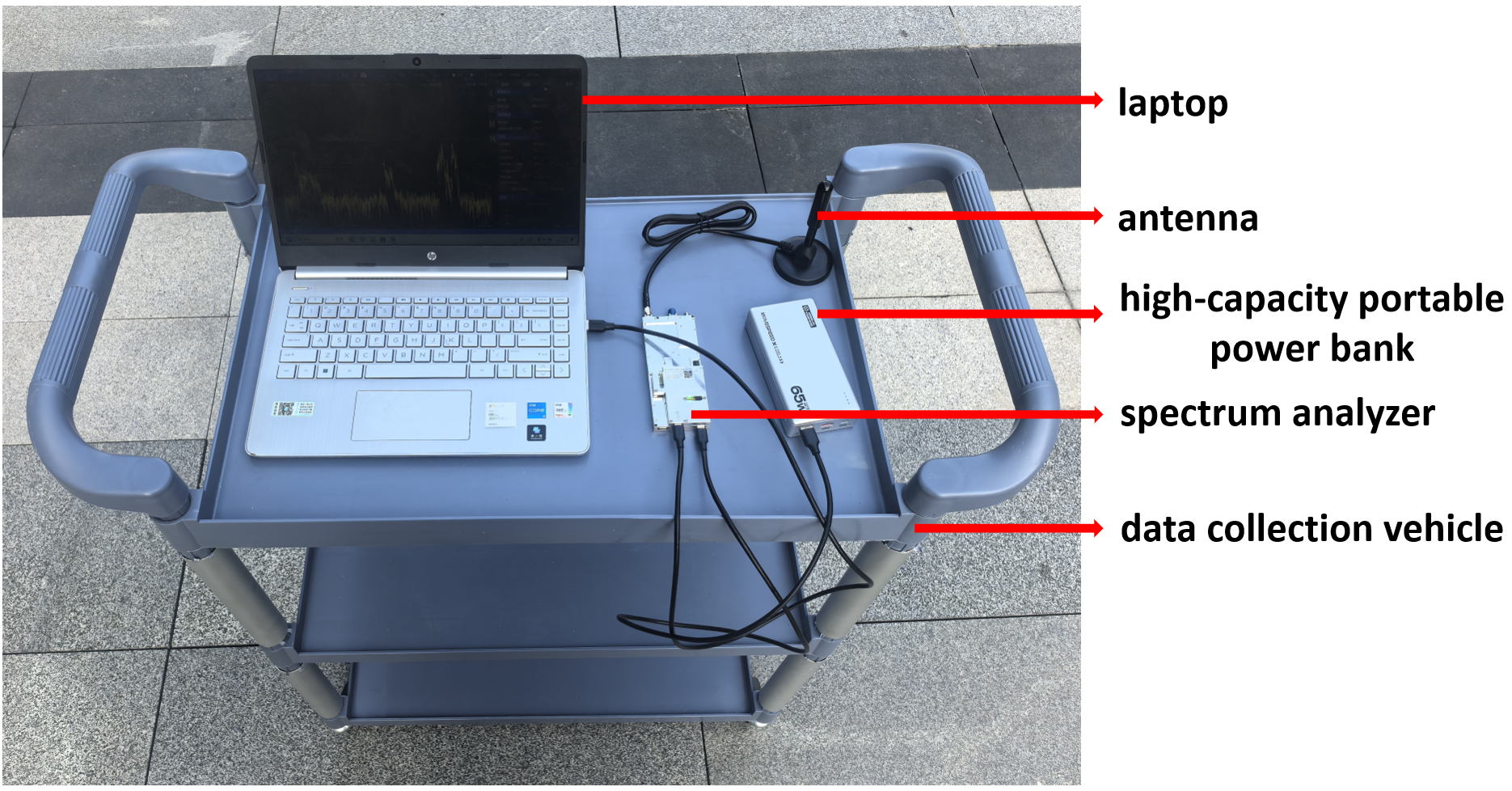}
    \caption{Data collection hardware system.}
    \label{fig:enter-label}
\end{figure}

\begin{figure}[!t]
    \centering
    \includegraphics[width=1\linewidth]{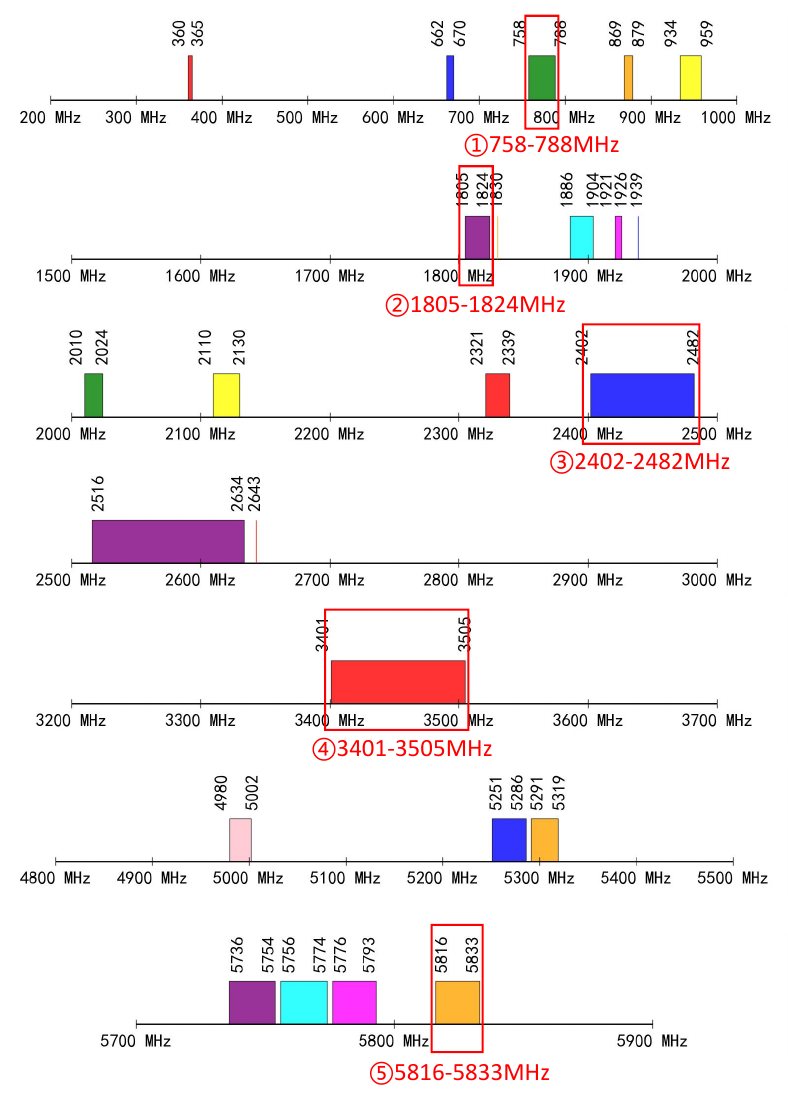}
    \caption{Visualization of environmental signal bands.}
\end{figure}

\begin{figure}[!t]
    \centering
    \includegraphics[width=1\linewidth]{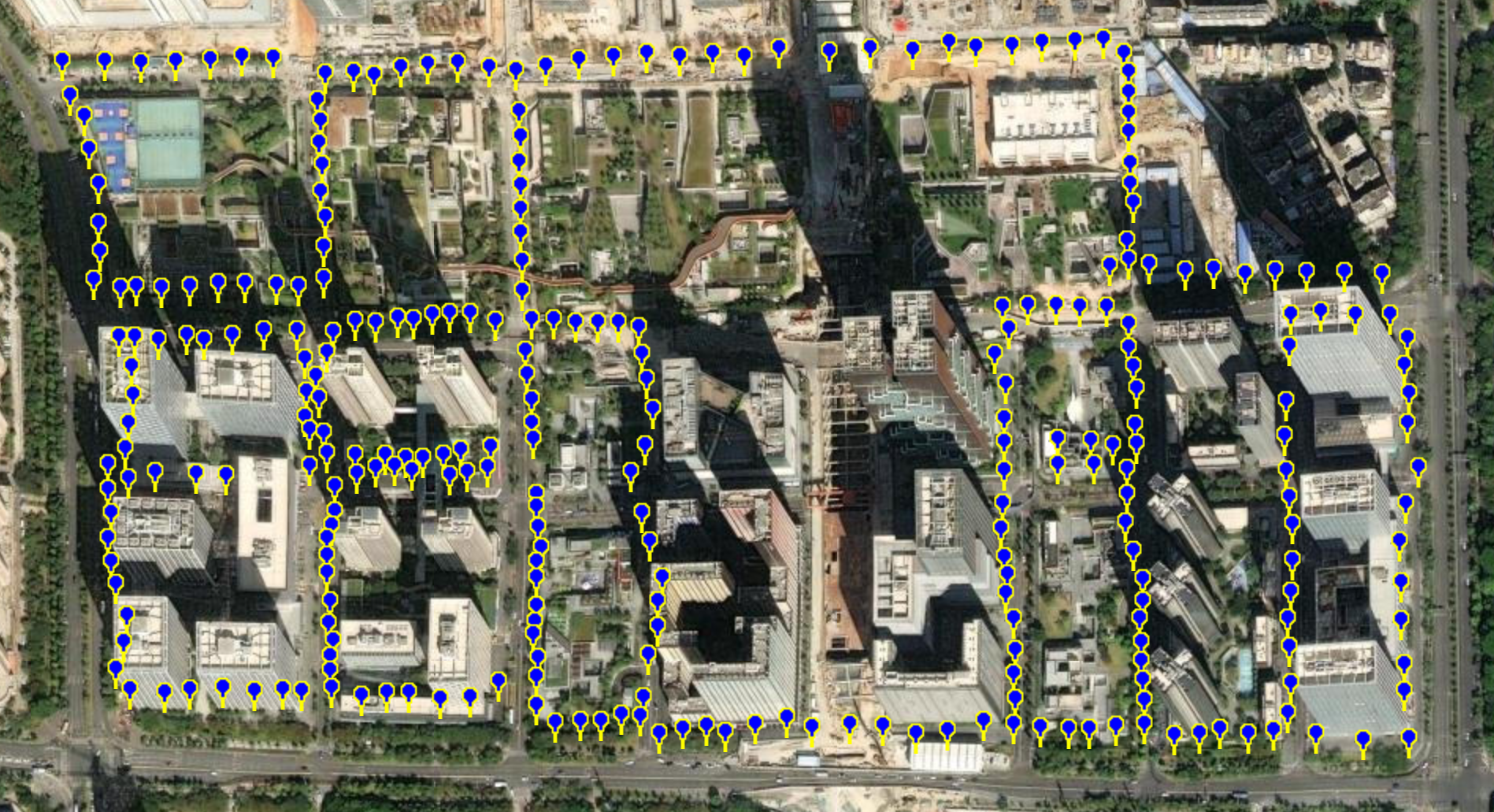}
    \caption{Illustration for sampling points in an urban area.}
    \label{fig:enter-label}
\end{figure}

\begin{figure*}[!t]
    \centering
    \includegraphics[width=1\linewidth]{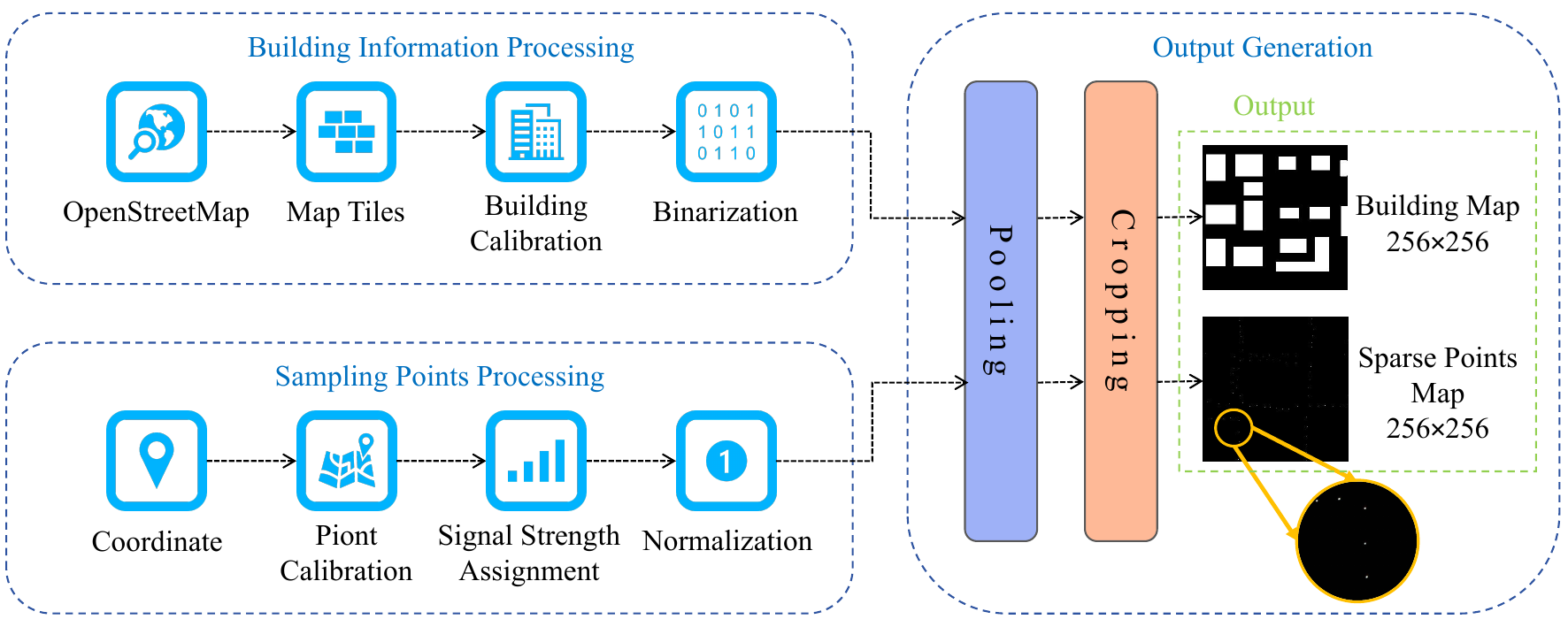}
    \caption{Real-world data processing pipeline.}
    \label{fig:enter-label}
\end{figure*}

\subsection{Measurement Settings}

In order to determine the measuring frequency bands, it is necessary to measure the environmental signal frequency bands first. We configure the spectrum analyzer to operate in real-time spectrum analysis mode, with a start frequency of 200 MHz and an end frequency of 6 GHz, and perform band-by-band scanning of the spectrum. Fig. 2 shows the existence of measured signals within the scanning frequency range. Based on the spectrum axis, five representative frequency bands commonly used for wireless signals are selected for the real-world radio map dataset: 758–788 MHz, 1805–1824 MHz, 2402–2482 MHz, 3401–3505 MHz, and 5816–5833 MHz. We take the average signal strength within each frequency band as the representative value. We collect spectrum data from three representative environments: urban, suburban, and campus areas. For each area, we gather signal measurements at 300 locations and record the geographic coordinates of each point. In total, 4,500 spectrum measurements are collected across 900 locations. Fig. 3 illustrates the sampling points in an urban area.

\subsection{Real-world Data Processing}

As shown in Fig. 4, in order to convert the collected real-world data into radio maps, the collected raw data is processed by the following steps:

\begin{enumerate}
\renewcommand{\labelenumi}{\arabic{enumi})}
\item \emph{Building Information Processing:} To characterize the influence of buildings on radio map construction, it is necessary to incorporate real-world building information into the mapping process. We download the map tiles corresponding to the measured area from OpenStreetMap~\cite{14}, record the latitude and longitude of its geographic boundaries, and mark the locations and shapes of the buildings. Then we obtain the building information with binary variables.

\item \emph{Sampling Points Processing:} We calibrate the sampling points according to the latitude and longitude information. Then we assign the signal strength values recorded by the spectrometer to the corresponding pixels of sampling points in the radio map, and normalize all the values.

\item \emph{Output Generation:} We eliminate the scale difference of map tiles in different regions by pooling operation, and then randomly crop them into small patches of $256 \times 256$ pixel size to form the model input.
\end{enumerate}

At this point, the construction of the real-world radio map dataset is completed. The dataset contains binarized building maps and normalized sparse points maps, each with a size of $256 \times 256$ pixels.

\section{RadioPiT: Model Design}

In this section, we propose RadioPiT, a PiT-based model enhanced by TTA strategy, which achieves high-accuracy radio map generation on ultra-sparse real-world data.

\subsection{Overview of PiT and TTA}

Before introducing our proposed model, we first provide a brief overview of the PiT model and the TTA strategy.

\paragraph{PiT Model} PiT is a generative architecture that integrates the Transformer attention mechanism~\cite{15}, enabling per-pixel generation. It excels at global modeling: its self-attention mechanism can capture long-range dependencies between arbitrary pixels. The attention mechanism enables the model to selectively focus on different parts of the input, assigning different weights to each part to emphasize the information most critical to the task. In the PiT model, each pixel is treated as a token, and its value is dynamically inferred from existing pixels, supporting per-pixel generation and local inpainting. This makes PiT particularly suitable for tasks involving incomplete data or sparse measurements, such as radio map generation.

\begin{figure*}[!t]
    \centering
    \includegraphics[width=1\linewidth]{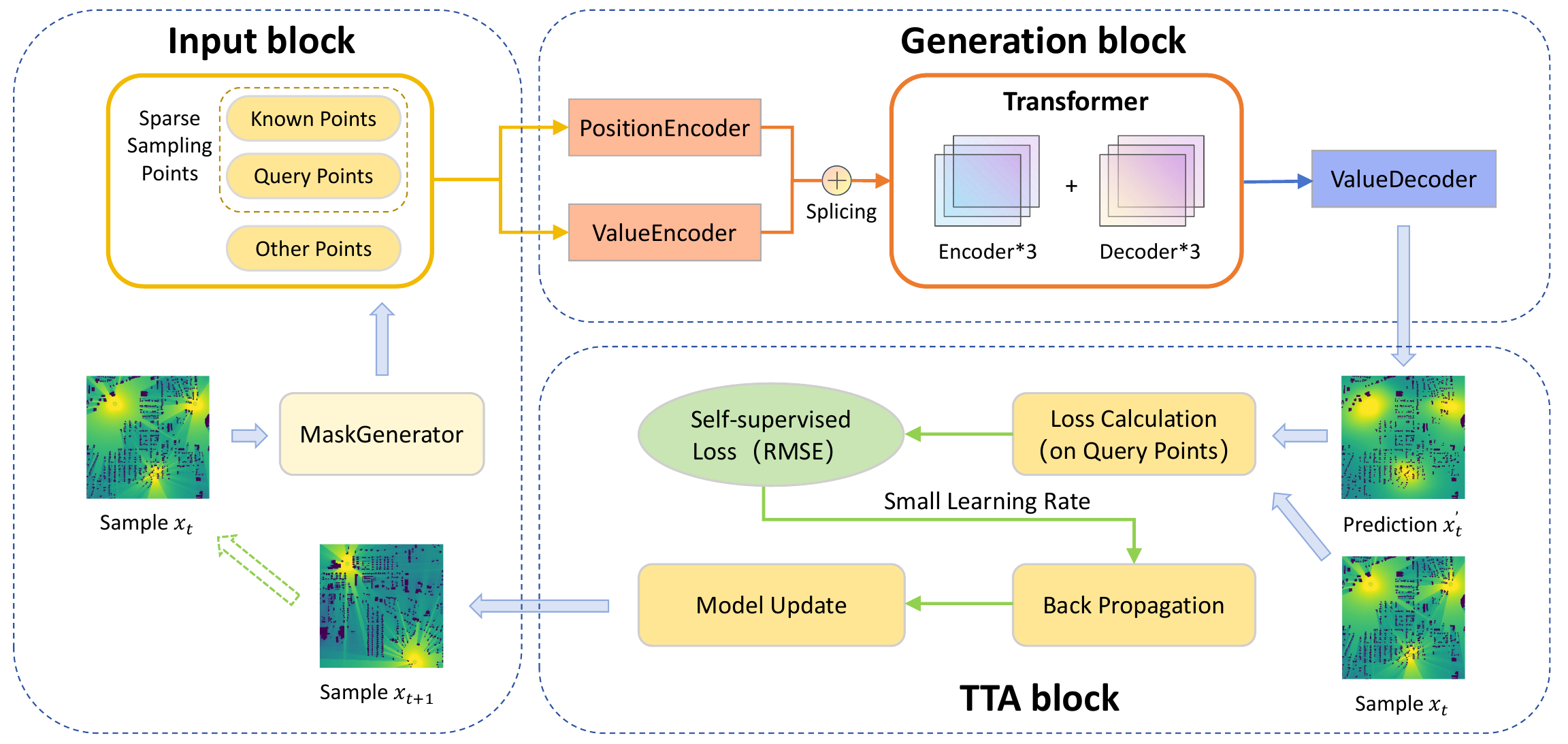}
    \caption{Architecture of the RadioPiT model.}
    \label{fig:enter-label}
\end{figure*}

\paragraph{TTA Strategy} TTA is a technique for online fine-tuning or adaptation of trained networks during the model testing phase. It is mainly applied in scenarios where the performance of the model declines due to changes in data characteristics when encountering new environments. It improves the prediction performance of the model by dynamically adjusting the model parameters according to each newly arrived sample during the test phase. Specifically, for each test sample, TTA first performs a forward propagation to obtain the output of the current sample. It then calculates the supervised loss between output results and true values, and performs a gradient update on the model. The learning rate in this process is generally extremely small (e.g., $5 \times 10^{-6}$), allowing only minor updates to the model parameters so as to avoid excessive influence of each test sample on the pre-trained model. The model with updated parameters is then used for predicting the next sample. Through this process, TTA gradually corrects the model bias during testing, thereby improving the prediction accuracy and robustness when processing samples with distributional discrepancies from training data.

\subsection{Framework of RadioPiT Model}

The architecture of the proposed RadioPiT model is shown in Fig. 5. RadioPiT consists of three blocks: an input block, a generation block, and a TTA block. 

\paragraph{Input Block}

In this block, sparse sampling points are extracted from the radio maps to construct the model input. The \textit{MaskGenerator} randomly selects a few sampling points from each radio map. These sampling points are split into known points and query points. Known points provide both coordinates and signal strength values to the model, whereas query points provide only coordinates and require the model to predict the values of their signal strength. The normalized coordinates and signal strength values of the known points are then fed into the generation block as prior information.

\paragraph{Generation Block}

In this block, the features of input data are processed by the PiT-based model to generate a high precision radio map of $256 \times 256$ pixels. The block receives the coordinates and signal strength values of the points from the input block, which are processed by the \textit{PositionEncoder} and the \textit{ValueEncoder} respectively, yielding high-dimensional position and value embeddings (with zero embedded to the positions of generated pixels). These two embeddings are then concatenated to form a unified embedding, which is then fed into the Transformer comprising three encoder layers and three decoder layers. The encoder layers employ multi-head self-attention~\cite{16} to capture the global spatial and spectral relationships among the known points. The decoder layers use cross-attention~\cite{17} to project the encoder’s global representations onto each point as needed. Finally, the decoder output is passed through the \textit{ValueDecoder} to produce the signal strength values of the entire radio map.

\paragraph{TTA Block}

This block implements TTA to enhance model performance. For each test sample, we obtain the predicted signal strength values at the query points from the complete radio map generated by the generation block. Then, the self-supervised loss is constructed from the RMSE between the true measured values and the model output of the query points. This process does not require any pseudo-labels or external annotations. We consider the following loss function:

\begin{equation}
\mathcal{L}_{\text{RMSE}}(X_i, \hat{X}_i) = \sqrt{ \frac{1}{n} \sum_{i=1}^{n} (X_i - \hat{X}_i)^2 },
\end{equation}

\noindent where $n$ is the number of query points, $X_i$ is the true value of the query point, and $\hat{X}_i$ is the value predicted by the model. Finally, the model performs a back propagation on this loss at an extremely small learning rate $\eta$ and applies a single gradient update step. The updated model is then employed for the radio map generation of the next sample. The model parameters $\theta_{t+1}$ used for predicting the test sample $\mathbf{x}_{t+1}$ can be written as:

\begin{equation}
\theta_{t+1} = \theta_t - \eta \cdot \nabla_{\theta_t} \mathcal{L}_{\text{RMSE}}(X_i, \hat{X}_i),
\end{equation}

\noindent where $\theta_t$ is the parameters used for predicting the test sample $\mathbf{x}_{t}$, and   $\nabla_{\theta_t} \mathcal{L}_{\text{RMSE}}(X_i, \hat{X}_i)$ is the gradient of the loss function with respect to the current model parameters $\theta_t$. 

With the above-mentioned three blocks, RadioPiT is able to perform radio map generation on ultra-sparse real-world data with high accuracy.

\section{Evaluation}

In this section, we conduct an experimental comparison between the proposed RadioPiT and baseline radio map generation methods, to demonstrate the superior performance of the RadioPiT model on radio map generation using ultra-sparse real-world data.

\subsection{Experiment Settings}

Due to the limited scale of the real-world radio map dataset, the RadioPiT model is first pre-trained on the BART-Lab Radiomap Dataset~\cite{18}, which is a simulation dataset generated with Altair WinProp software. It covers frequency bands of 1750 MHz, 2750 MHz, 3750 MHz, 4750 MHz, and 5750 MHz. Each band contains 2000 coarse-resolution radio maps of approximately $500 \times 500$ pixels. We divide these radio maps into training and validation sets at a 4:1 ratio. All the radio maps are randomly cropped to $256 \times 256$ pixels before being sent to the model. During the training process, \textit{MaskGenerator} randomly samples 50 known points in each radio map, which are fed into the model to generate the signal strength values of 1500 query points and other remaining points. The RMSE between the predicted and real values at the query points is used as the loss for model training. The pre-training phase is performed using an AdamW optimizer in conjunction with a cosine annealing learning rate scheduler, with an initial learning rate of $10^{-4}$ and a weight decay of $10^{-4}$. The batch size is set to 64 and the number of training epochs is 200. 

After obtaining the RadioPiT pre-trained on the simulation radio map dataset, we fine-tune and test the model using the real-world radio map dataset. Each radio map contains 10 to 100 sampling points. These points are randomly divided into known points and query points in the ratio of 2:1. The model input only includes the coordinates and signal strength values of the known points, along with the coordinates of the query points and all other remaining points. The model outputs the predicted signal strength values at each point and then performs TTA to update its parameters accordingly. TTA is conducted using an Adam optimizer with a learning rate of $5 \times 10^{-6}$. The updated model is subsequently employed for the inference of the next sample. 

In the experiments, we select the Kriging method, a representative of interpolation method, and RadioUNet, a representative GAI model, as baseline methods for comparison.

\begin{itemize}
\item Kriging~\cite{19}: An optimal geostatistical method that estimates signal strength at unknown locations based on known measurements and spatial autocorrelation.

\item RadioUNet~\cite{8}: A deep learning model based on the U-Net encoder–decoder architecture, specifically designed for radio map prediction.
\end{itemize}

\subsection{Experimental Results}

Fig. 6 shows the comparison of radio map generation performance across Kriging, RadioUNet, and RadioPiT on simulation and real-world radio map datasets. RadioUNet and RadioPiT are pre-trained on the same simulation radio map dataset, the BART-Lab Radiomap Dataset. RadioUNet exhibits notable performance degradation when applied to the real-world radio map dataset, even performing worse than Kriging. This is primarily due to the significant distributional discrepancies between simulation and real-world data, which make it difficult for the pre-trained parameters of RadioUNet to adapt to the real-world data. By contrast, our proposed RadioPiT maintains stable and superior performance on the real-world radio map dataset due to the application of the TTA strategy, which mitigates the adverse effects of such distributional discrepancies. Compared to Kriging and RadioUNet, the RMSE of our RadioPiT decreases 15.2\% and 21.9\% respectively, indicating the advantages of RadioPiT for radio map generation on ultra-sparse real-world data. 

\begin{figure}[h]
    \centering
    \includegraphics[width=1\linewidth]{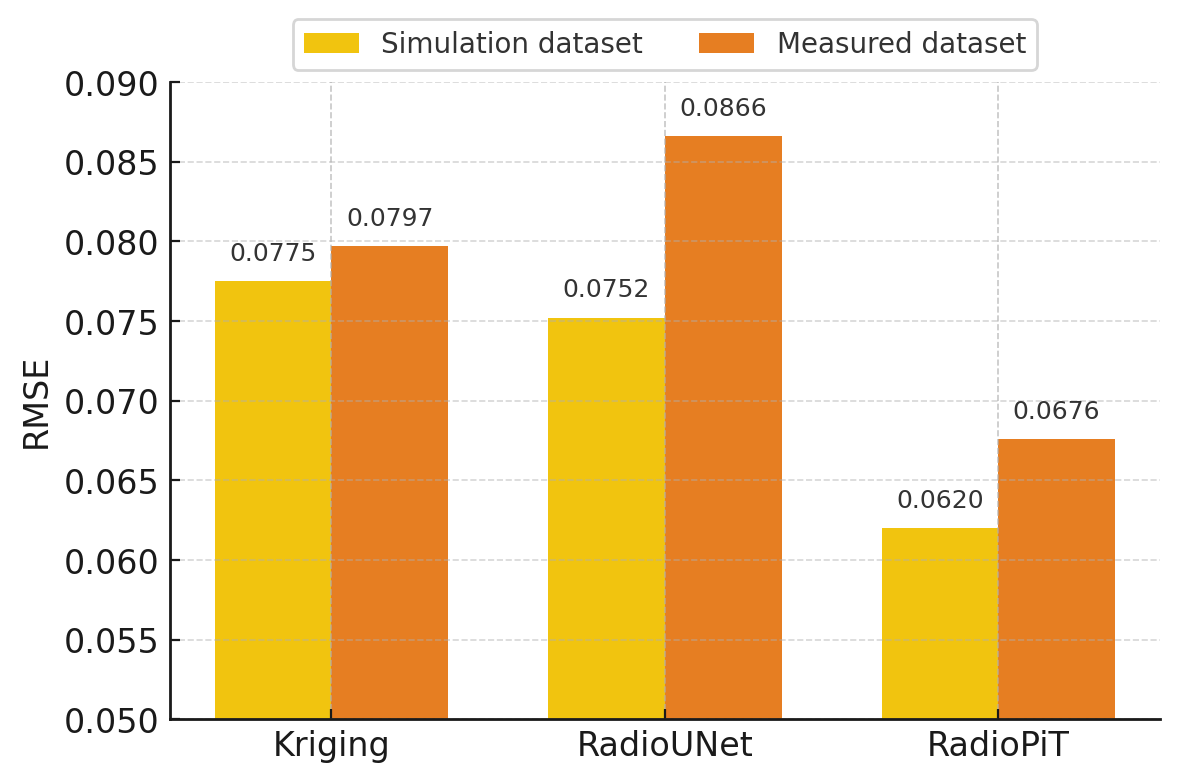}
    \caption{Comparison of radio map generation performance across different methods on simulation and real-world radio map datasets.}
    \label{fig:enter-label}
\end{figure}

Fig. 7 presents the results of the ablation study, showcasing the effects of the TTA strategy and the PiT architecture on model performance. It can be seen that the RMSE of RadioPiT on the real-world radio map dataset reaches 0.0741 when TTA is not applied, representing a 9.6\% RMSE increment compared to the TTA-enhanced version. Under the same TTA strategy, RadioUNet, which lacks the PiT architecture, yields an RMSE of 0.0770 on the real-world radio map dataset, indicating a 13.9\% RMSE increment compared to RadioPiT.

\begin{figure}[h]
    \centering
    \includegraphics[width=1\linewidth]{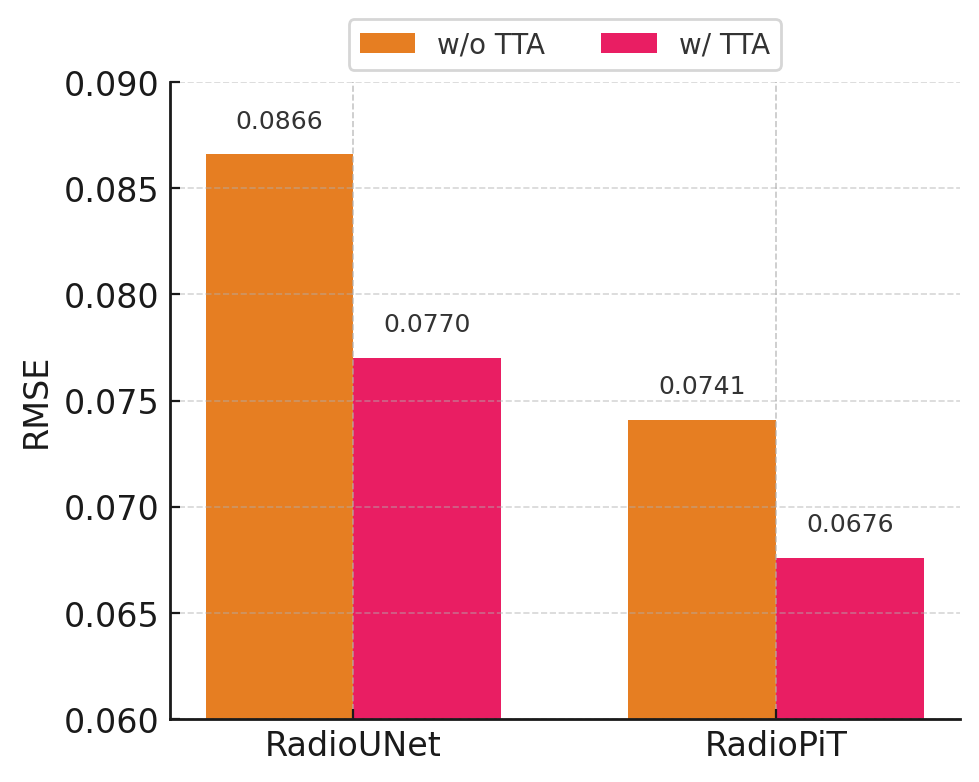}
    \caption{Ablation study on the effects of the TTA and the PiT architecture.}
    \label{fig:enter-label}
\end{figure}

\begin{figure*}[!t]
    \centering
    \includegraphics[width=0.99\linewidth]{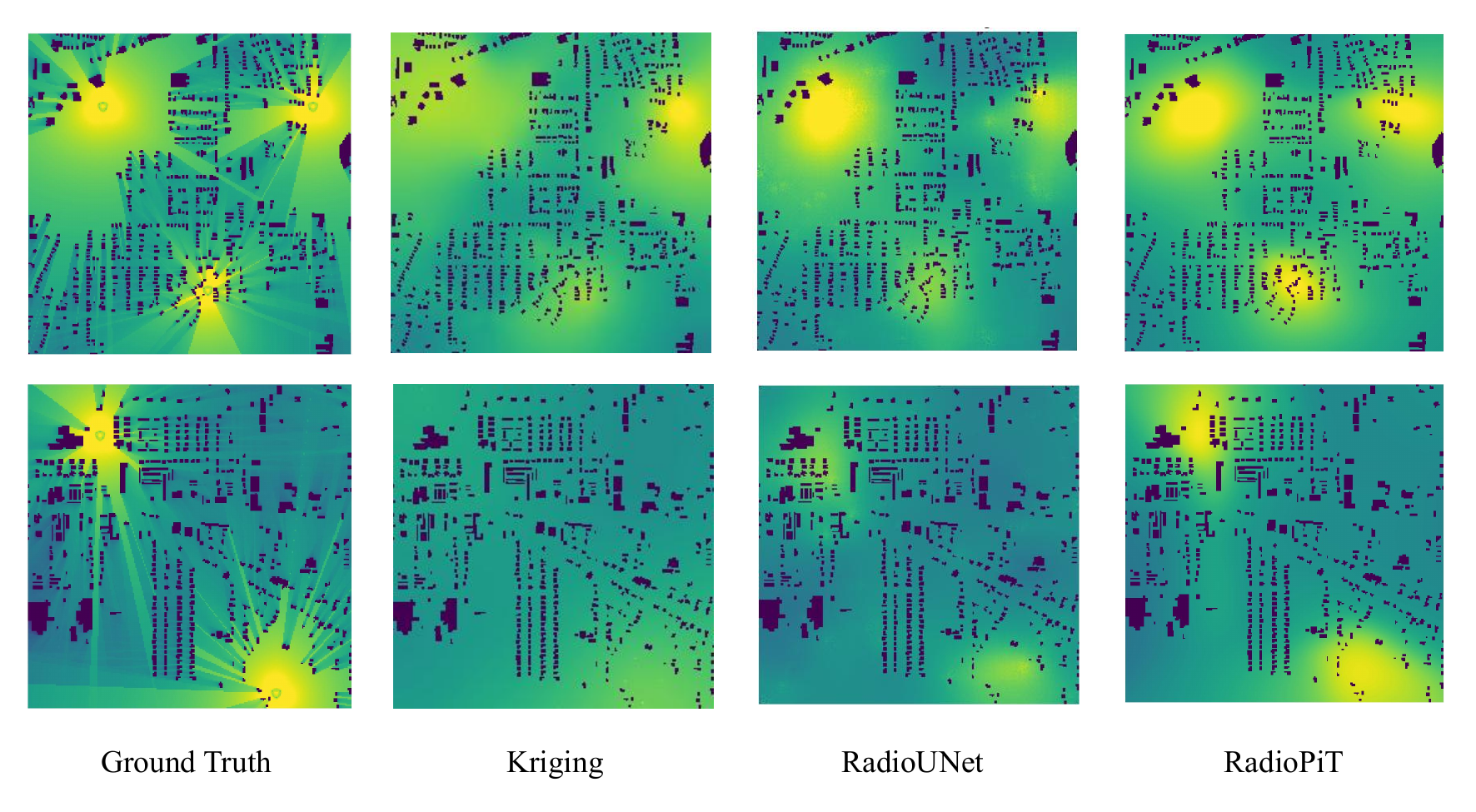}
    \caption{Radio map generation performance of different methods.}
    \label{fig:enter-label}
\end{figure*}

Fig. 8 illustrates the radio map generation performance of different methods in the 1805–1824 MHz frequency band. It can be intuitively observed that under ultra-sparse data conditions, both Kriging and RadioUNet exhibit substantial deviations from the ground-truth values, especially near the transmitters. In contrast, RadioPiT achieves superior radio map generation performance in such sparse settings, benefiting from the PiT architecture equipped with the attention-enabled per-pixel generation mechanism and the application of the TTA strategy.

\section{Conclusion}

In this paper, we have constructed a real-world radio map dataset with a self-developed measurement system, and have proposed a new PiT-based model enhanced with the TTA strategy, named RadioPiT, which enables radio map generation on ultra-sparse real-world data. Experimental results have demonstrated that our proposed RadioPiT outperforms baseline methods such as Kriging and RadioUNet, with RMSE decrement of 15.2\% and 21.9\%, respectively. This performance gain is attributed to the use of the PiT architecture with the attention-enabled per-pixel generation mechanism, and the TTA strategy which alleviates the negative effects caused by the distributional discrepancies between simulation and real-world data.

\section*{Acknowledgment}

This work was supported in part by the Key Project of Beijing Natural Science Foundation under Grant L253004; in part by the National Science Foundation under Grant 62401302; in part by the Guangdong Basic and Applied Basic Research Foundation under Grant 2023B0303000019; in part by the Shenzhen Science and Technology Program under Grant JCYJ20241202125910015; in part by the Beijing Outstanding Young Scientist Program JWZ020240102001; in part by the National Science Foundation under Grant 62401025.

\end{document}